\documentclass[iop,apjl]{emulateapj}
\usepackage{xcolor} 

\usepackage{apjfonts}

\newcommand\ergs{erg~s$^{-1}$}

\newcommand{\nar}{New A Rev.}

\begin{document}

\title{Nature of the soft ULX in NGC 247: super-Eddington outflow and transition between the supersoft and soft ultraluminous regimes}

 \author{
 Hua Feng\altaffilmark{1}, 
 Lian Tao\altaffilmark{2}, 
 Philip Kaaret\altaffilmark{3}, 
 Fabien Gris\'{e}\altaffilmark{4}}

\altaffiltext{1}{Department of Engineering Physics and Center for Astrophysics, Tsinghua University, Beijing 100084, China}
\altaffiltext{2}{Cahill Center for Astronomy and Astrophysics, California Institute of Technology, Pasadena, CA 91125, USA}
\altaffiltext{3}{Department of Physics and Astronomy, University of Iowa, Iowa City, IA 52242, USA}
\altaffiltext{4}{Observatoire astronomique de Strasbourg, Universit\'{e} Strasbourg, CNRS, UMR 7550, 11 rue de l'Universit\'{e}, F-67000 Strasbourg, France}

\shorttitle{The nature of NGC 247 ULX}
\shortauthors{Feng et al.}

\begin{abstract}

We report on {\it XMM-Newton}/{\it Chandra}/{\it Swift}/{\it HST} observations of the ultraluminous X-ray source (ULX) in NGC 247, which is found to make transitions between the supersoft ultraluminous (SSUL) regime with a spectrum dominated by a cool ($\sim 0.1$~keV) blackbody component and the soft ultraluminous (SUL) regime with comparable luminosities shared by the blackbody and power-law components. Multi-epoch observations revealed an anti-correlation between the blackbody radius and temperature, $R_{\rm bb} \propto T_{\rm bb}^{-2.8 \pm 0.3}$, ruling out a standard accretion disk as the origin of the soft X-ray emission.  The soft X-ray emission is much more variable on both short and long timescales in the SSUL regime than in the SUL regime. We suggest that the SSUL regime may be an extension of the ultraluminous state toward the high accretion end, being an extreme case of the SUL regime, with the blackbody emission arising from the photosphere of thick outflows and the hard X-rays being emission leaked from the embedded accretion disk via the central low-density funnel or advected through the wind. However, the scenario that the supersoft ULXs are standard ULXs viewed nearly edge-on cannot be ruled out. Flux dips on a timescale of 200~s were observed. The dips cannot be explained by an increase of absorption, but could be due to the change of accretion rate or related to thermal fluctuations in the wind or disk. The optical emission of NGC 247 ULX exhibits a blackbody spectrum at a temperature of 19,000~K with a radius of 20 $R_\sun$, likely arising from an OB supergiant companion star.

\end{abstract}


\section{Introduction}
\label{sec:intro}

Ultraluminous X-ray sources (ULXs) are non-nuclear accreting X-ray sources whose apparent luminosity exceeds the Eddington limit of Galactic stellar mass black holes ($3 \times 10^{39}$~\ergs\ for a 20 $M_\sun$ black hole).  Since they were discovered in 1970s \citep{Fabbiano1989}, their physical nature is still unclear \citep[for a review see][]{Feng2011}. Deep X-ray studies in recent years favor the interpretation that most ULXs are likely due to supercritical accretion onto stellar mass black holes \citep[e.g.,][]{Gladstone2009,Middleton2012,Sutton2013,Middleton2015}, with dynamical evidence in a couple cases \citep{Liu2013,Motch2014}. Theoretical models and numerical simulations \citep{King2003,Poutanen2007,Ohsuga2011,Dotan2011,Kawashima2012,Jiang2014,Shen2015} suggest that massive outflows are inevitable for supercritical accretion. The outflows may be optically-thick and play an important role in shaping the emergent X-ray energy spectrum and timing behavoir. It is generally agreed that the disk will be thick and inefficient in radiation, e.g., the slim disk model \citep{Abramowicz1988}. However, recent 3D radiation magneto-hydrodynamic simulations of supercritical accretion by \citet{Jiang2014} show that vertical advection induced by turbulence generated by the magneto-rotational instability can produce radiative efficiency as high as standard thin-disk accretion and nearly isotropic emission. However, the ULX population is heterogeneous.  Extreme ULXs like ESO 243-49 HLX-1 and M82 X-1 are good candidates for intermediate mass black holes \citep{Farrell2009,Servillat2011,Feng2010,Pasham2014}.  While a transient ULX in M82 with a peak luminosity over $10^{40}$~\ergs\ \citep{Feng2007_M82,Brightman2016} is powered by accretion onto a neutron star \citep{Bachetti2014}.  The nature of ULXs is still a question under debate and more observations are needed. 

Unlike typical ULXs which exhibit substantial emission above 2 keV, there exists a sub-class whose X-ray spectrum is dominated by emission from a blackbody-like ($kT_{\rm bb} \approx 0.05-0.2$~keV) component below 2 keV \citep{DiStefano2003,DiStefano2004}. Well studied examples include those in M51 \citep{DiStefano2003,Terashima2004,Dewangan2005,Terashima2006}, M81 \citep{Swartz2002,Swartz2003,Liu2008,LiuDiStefano2008,Liu2015}, M101 \citep{Pence2001,Mukai2003,Kong2004,Jenkins2004,Kong2005,Mukai2005,Liu2009,Liu2013,Shen2015,Soria2016}, NGC 247 \citep{Winter2006,Jin2011,Tao2012_NGC247}, NGC 300 \citep{Read2001,Kong2003},  the Antennae \citep[NGC4038/4039;][]{Fabbiano2003}, and NGC 4631 \citep{Carpano2007,Soria2009}.  Multi-epoch observations of the canonical soft ULX in M101 revealed dramatic variability in the observed X-ray flux by 2--3 orders of magnitude \citep{Soria2016}. Such strong variability is rare for broad-band ULXs. 

\citet{Urquhart2016} performed a comprehensive X-ray study of the brightest soft ULXs. They found a correlation between the blackbody temperature and emission radius, $R_{\rm bb} \propto T_{\rm bb}^{-2.2 \pm 0.5}$, for their sample.  \citet{Soria2016} identified a similar relation, roughly $R_{\rm bb} \propto T_{\rm bb}^{-2}$, for the soft ULX in M101 from multi-epoch observations. This seems to be a universal relation for soft ULXs.  It has been proposed that the soft spectrum of soft ULXs could be emission from the photosphere of optically thick outflows, which obscure or soften the hard emission from the inner disk \citep{Feng2011,Shen2015,Soria2016,King2016}. \citet{King2016} suggest that the soft ULXs and broad-band ULXs are the same type of objects viewed at different angles. \citet{Urquhart2016} suggest that the soft ULXs could be a special state of the normal ULXs; they may transition to be broad-band ULXs when the photosphere becomes smaller and hotter ($kT_{\rm bb} \sim 150$~eV). These scenarios are yet to be tested, as no source has been observed to date to transition between the two states or regimes. 

In this paper, we report multiwavelength observations of one of the brightest soft ULXs in NGC 247 (R.A. = 00$^{\rm h}$47$^{\rm m}$03\fs88, decl. =  $-$20\arcdeg47\arcmin44\farcs3, J2000.0). It was first tagged as a ULX from a short and badly contaminated {\it XMM-Newton} observation \citep{Winter2006}.  Later, its soft nature was secured via a second {\it XMM-Newton} observation \citep{Jin2011}. The optical counterpart to the X-ray source was identified using aligned {\it Chandra} and {\it Hubble Space Telescope} ({\it HST}) images \citep{Tao2012_NGC247}.  Here, we attempt to unveil the physical origin of the optical emission using multi-band {\it HST} observations and the nature of the X-ray emission using multi-epoch observations with {\it XMM-Newton}, {\it Chandra}, and {\it Swift}. We adopt a distance of 3.4~Mpc estimated using IR Cepheids to the galaxy NGC 247 \citep{Gieren2009}.

We use the term {\it soft} instead of  {\it supersoft} to refer to these sources, in particular for NGC 247 ULX, because they are always soft but not always supersoft. They vary in temperature and spectral shape, and some of them in some occasions may display substantial emission between 1 and 2~keV or even above 2~keV. In their hardest state, they may not be classified as supersoft sources based on the criteria defined by \citet{DiStefano2003} due to the presence of a relatively strong power-law component.

\section{Observations and data reduction}
\label{sec:obs}

\tabletypesize{\scriptsize}
\begin{deluxetable}{ccccc}
\tablecolumns{5}
\tablewidth{\linewidth}
\tablecaption{Observations used in the paper
\label{tab:obs}}
\tablehead{
\colhead{Telescope} & \colhead{ID} & \colhead{Date} & \colhead{Exposure} & \colhead{Detector/Filter}
}
\startdata
{\it XMM} & 0601010101 & 2009 Dec 27--28 & 30 ks & MOS1/MOS2/PN  \\
{\it XMM} & 0728190101 & 2014 Jul 01 & 30 ks & MOS1/MOS2/PN  \\
{\it Chandra} & 12437 & 2011 Feb 01 & 5.0 ks & ACIS-S \\
{\it Chandra} & 17547 & 2014 Nov 12 & 5.0 ks & ACIS-S \\
{\it Swift} & 00033469 & 2014 Oct--2015 Nov & 28 $\times$ 2 ks & XRT \\
{\it HST} & ic8ea1010 & 2014 Jun 30 & 468 s & WFC3/UVIS/F225W \\
{\it HST} & ic8ea1020 & 2014 Jun 30 & 466 s & WFC3/UVIS/F336W \\
{\it HST} & ic8ea1030 & 2014 Jun 30 & 96 s & WFC3/UVIS/F438W \\
{\it HST} & ic8ea1040 & 2014 Jun 30 & 96 s & WFC3/UVIS/F606W \\
{\it HST} & ic8ea1050 & 2014 Jun 30 & 200 s & WFC3/UVIS/F814W \\
{\it HST} & ic8ea1060 & 2014 Jun 30 & 257 s & WFC3/IR/F105W \\
{\it HST} & ic8ea1070 & 2014 Jun 30 & 303 s & WFC3/IR/F160W \\
{\it HST} & jc8e01020 & 2014 Jun 30 & 2348 s & ACS/SBC/PR130L 
\enddata
\end{deluxetable}

The observations used in this paper are listed in Table~\ref{tab:obs}. NGC 247 was observed with {\it XMM-Newton} on 2001 July 08 (ObsID 0110990301), 2009 December 07 (ObsID 0601010101), and 2014 July 01 (ObsID 0728190101). The first observation, from which the source was found to exhibit a soft spectrum \citep{Winter2006,Jin2011}, was short and strongly contaminated by background flares and is not considered for spectral analysis here. The second observation has been previously described by \citet{Jin2011}, who confirmed that the source spectrum consists of a dominant soft, thermal component and a weak power-law component, reminiscent of the thermal state spectrum of Galactic black hole binaries. The third observation was obtained as of our multiwavelength campaign with {\it HST} and is described in detail below. In addition to the sensitive {\it XMM-Newton} observations, we also report results from 2 short {\it Chandra} observations (5 ks each) and 28 {\it Swift} snapshots (2 ks each) of the source. 

\subsection{{\it XMM-Newton}}

New event files were produced with updated calibration files.  Clean exposures were selected from time intervals where the background flux is within $\pm3\sigma$ of the mean quiescent level. For the 2009 observation (ObsID 0601010101), the total clean exposure is 22.3~ks for PN, 29.0~ks for MOS1, and 30.2~ks for MOS2. For the 2014 observation (ObsID 0728190101), the exposure is 27.2~ks for PN, 31.0~ks for MOS1, and 32.4~ks for MOS2. 

\begin{figure}
\centering
\includegraphics[width=\columnwidth]{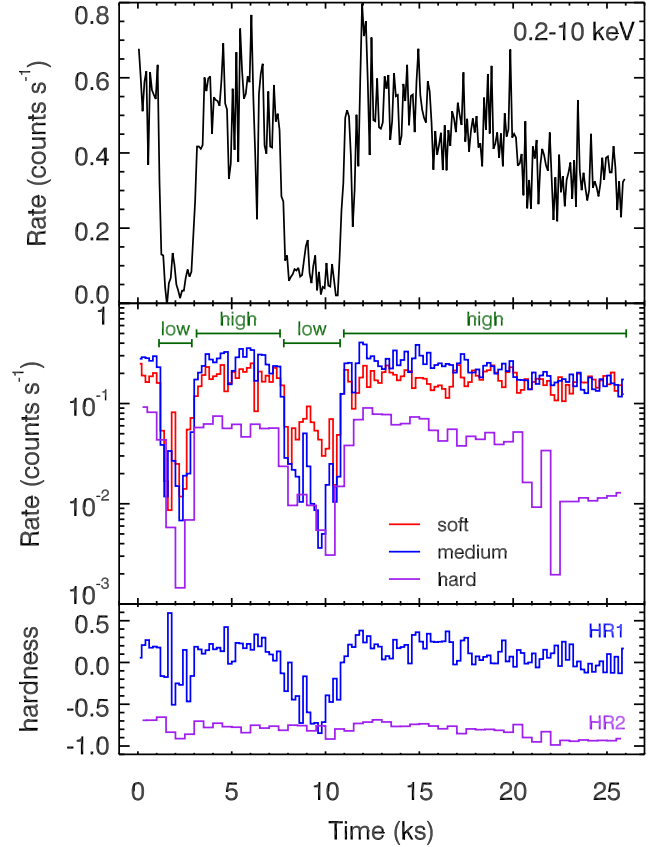}
\caption{{\it XMM-Newton} PN lightcurves of NGC 247 ULX from the 2014 observation in the longest continuous and clean time interval with an exposure of 26.1~ks. {\bf Top}: lightcurve in the full band (0.2--10 keV). {\bf Middle}: lightcurves in the soft band (0.2--0.75 keV), medium band (0.75--1.5 keV), and hard band (1.5--10 keV). {\bf Bottom}: hardness ratios defined as  HR1 = (medium $-$ soft) $/$ (medium $+$ soft) and HR2 = (hard $-$ medium $-$ soft) $/$ (hard $+$ medium $+$ soft).  The lightcurve has a time step of 100~s for the full band, 200~s for the soft and medium bands, and 500~s for the hard band. 
\label{fig:lc}}
\end{figure}

\begin{figure*}[tbh]
\centering
\includegraphics[width=\textwidth]{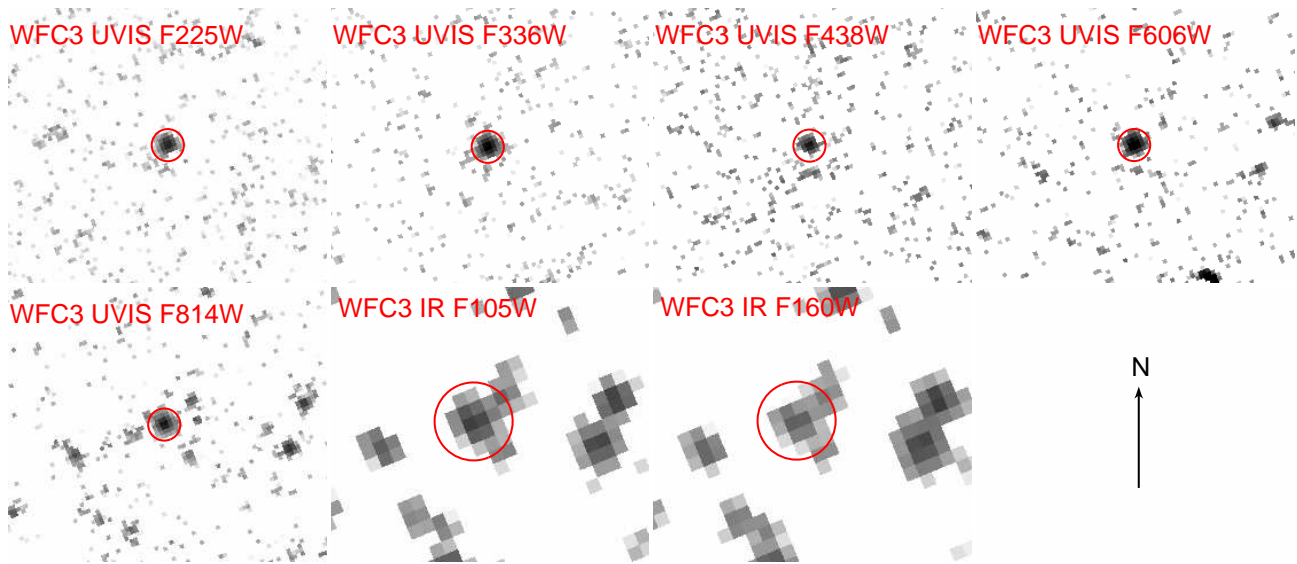}
\caption{{\it HST} WFC3 images around the ULX region.  The circles indicate the optical counterpart of the ULX. The arrow points north and has a length of 1\arcsec.
\label{fig:wfc3}}
\end{figure*}

Figure~\ref{fig:lc} shows PN lightcurves for the longest (26.1 ks) continuous and low background time interval from the 2014 observation.  Events were extracted from a 32\arcsec-radius circular region surrounding the source and selected to have FLAG = \#XMMEA\_EP and PATTERN $\le$ 4.  The background contribution was estimated from several nearby regions and subtracted. Lightcurves are displayed in the full band of 0.2--10 keV, a soft band of 0.2--0.75~keV, a medium band of 0.75--1.5~keV, and a hard band of 1.5--10~keV. The bottom panel of Figure~\ref{fig:lc} shows hardness ratios, defined as HR1 = (medium $-$ soft) $/$ (medium $+$ soft) and HR2 = (hard $-$ medium $-$ soft) $/$ (hard $+$ medium $+$ soft).  HR1 is sensitive to the temperature of the soft thermal component and HR2 is sensitive to the flux ratio of the hard power-law component versus the soft thermal component. As one can see, the source displayed a dramatic change in flux (dips) and the medium-to-soft hardness ratio. Time intervals for the low- and high-flux states are marked in Figure~\ref{fig:lc}; their energy spectra were extracted and analyzed separately. 

The same aperture was used to extract the source spectra and a few nearby circular regions on the same CCD chip at similar readout distances but off the readout direction were used to extract the background spectra, both from events with FLAG $=$ 0 and PATTERN $\le$ 12 for MOS or $\le 4$ for PN. Each spectrum was grouped in the energy band from 0.2~keV to roughly 10~keV to have 4 bins per (FWHM) spectral resolution element or at least 25 counts. 

\subsection{{\it Chandra}}

New level 2 event files were created using the {\tt repro} script in CIAO. An elliptical source region found by the {\tt wavdetect} task was adopted for spectrum extraction, which was 1\farcs62 in semi-major radius and 1\farcs15 in semi-minor radius for the 2011 observation (ObsID 12437), and 1\farcs46 and 1\farcs32, respectively, for the 2014 observation (ObsID 17547). A nearby large, source-free region was used to extract a background spectrum.  For ObsID 12437, there are 160 counts detected in the energy band of 0.3--1 keV, 7 counts in 1--2 keV, and only 1 count in 2-8 keV. For ObsID 17547, there are 103 counts in 0.3--1 keV, 70 counts in 1--2 keV, and 4 counts in 2--8 keV. Thus, in order to perform $\chi^2$ fitting, we grouped the energy spectrum by at least 15 counts per bin from 0.3 to 1 keV for ObsID 12437, and in the same manner from 0.3 to 2 keV for ObsID 17547. This results in 8 energy bins for ObsID 12437 and 9 for ObsID 17547. 

\subsection{{\it Swift}}

The NGC 247 galaxy was monitored with the {\it Swift} X-ray telescope (XRT) from 2014 October 13 to 2015 November 19 under the program 00033469. The cadence was a week in the beginning (once interrupted due to sun avoidance) and changed to a month for the last six observations. The requested exposure for each observation was 2 ks, but it varied from 0.9 to 2.5 ks.   The detected photons in the source aperture ranges from 8 to 51 for a single observation, and the background contribution is 3\% typically and 11\% in the worst case.   The source displayed a correlation between count rate and hardness ratio in individual XRT observations.  So, in order to perform $\chi^2$ fitting, we combined events from observations with similar background corrected count rates.  We divided the 28 individual observations into 5 groups, each one with a total number of 120--130 source counts. We extracted the source spectrum from an aperture of 20 pixels in radius from the combined event files. The background spectrum was extracted from a nearby source-free region, verified using the deep {\it XMM-Newton} image. Each spectrum was grouped to have at least 15 counts per bin from 0.3 to 2 keV. 

\subsection{{\it HST}}

\tabletypesize{\footnotesize}
\begin{deluxetable}{cccl}
\tablecolumns{4}
\tablewidth{\columnwidth}
\tablecaption{Observed magnitudes of the ULX with {\it HST} WFC3.
\label{tab:mag}}
\tablehead{
\colhead{Channel} & \colhead{Filter} & \colhead{Pivot wavelength (\AA)} & \colhead{Vega mag}
}
\startdata
UVIS & F225W & 2359 & $20.77 \pm 0.03$ \\
UVIS & F336W & 3355 & $21.03 \pm 0.02$ \\
UVIS & F438W & 4325 & $22.47 \pm 0.05$ \\
UVIS & F606W & 5887 & $22.22 \pm 0.03$ \\
UVIS & F814W & 8024 & $22.07 \pm 0.03$ \\
IR & F105W    & 10552 & $21.67 \pm 0.02$ \\
IR & F160W    & 15369 & $21.60 \pm 0.03$ 
\enddata
\end{deluxetable}

The {\it HST} observations (proposal ID 13425) started about 18 hours before the beginning of the {\it XMM-Newton} observation and were finished in two orbits ($\sim$2.4 hours). Images were taken using the Wide Field Camera 3 (WFC3) with both the UVIS and IR Channels. A number of wide-band filters (see Table~\ref{tab:obs}) were used to cover a wavelength range from 1900 to 18000\AA.  A $512 \times 512$ subarray was adopted for both UVIS and IR to reduce the readout time. Images around the ULX region for different filters are shown in Figure~\ref{fig:wfc3}.  Point spread function (PSF) photometry was performed on the calibrated, flat-fielded images using the DOLPHOT 2.0 package \citep{Dolphin2000}. We first ran {\tt wfc3mask} to mask bad pixels, and then split UVIS images using {\tt splitgroups}.  After estimating the sky value with {\tt calcsky}, we ran {\tt dolphot} to get the aperture and charge transfer efficiency (CTE) corrected Vega magnitudes. We compared with aperture photometry and found well consistent results in the UVIS bands. For IR images, the ULX region is crowded with faint sources and the difference in PSF and aperture photometry is as large as 0.7 mag in the F160W band. We thus adopted results from the PSF photometry, which should produce more reliable results for sources in crowded regions, but note that the true errors for IR flux taking into account systematics in photometry may be larger. The SYNPHOT package was used for flux conversion.  The observed magnitudes in different bands are listed in Table~\ref{tab:mag}.

\begin{figure}
\centering
\includegraphics[width=0.8\columnwidth]{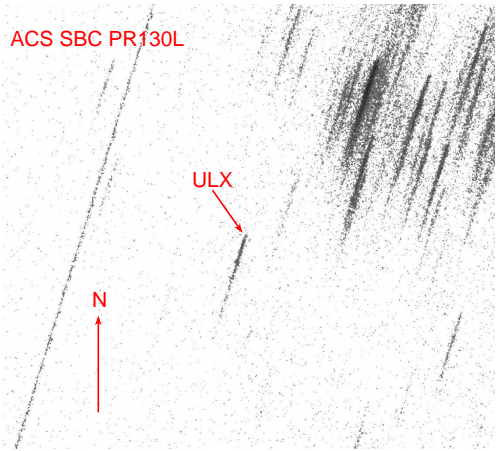}
\caption{The dispersed image from the grism ACS SBC PR130L around the ULX. The arrow points north and has a length of 3\arcsec. 
\label{fig:sbc}}
\end{figure}

The low resolution grism PR130L in the Solar Blind Channel (SBC) of the Advanced Camera for Surveys (ACS) was used to produce a slitless spectrum  in the UV band. The dispersed image of the PR130L prism (see Figure~\ref{fig:sbc}) consists of four sub-exposures. The last sub-exposure (jc8e01e3q) had a low signal-to-noise ratio (S/N) due to an increase of detector temperature and was not used. Using version 2.4.4 of the aXe software package in PyRAF/STSDAS, spectra were extracted with an aperture width of 12 pixels. The direct image, taken through the F165LP filter, was used to provide the wavelength zero point on the prism image. The spectrum covers the wavelength range from 1300\AA\ to 1900\AA. Due to the faintness of the object, no significant emission or absorption line features can be detected. The three spectra from the sub-exposures were stacked and binned to have a S/N of at least 15 per spectral bin (except for the last bin which has a S/N of 9.4). 

\section{Analysis and results}

\subsection{Spectral modeling to the X-ray data}

The {\it XMM-Newton} PN and MOS spectra from the 2014 observation were fitted jointly using XSPEC. A constant scaling factor on the MOS flux with respect to the PN flux is included for the high state spectra as a free parameter to account for possible difference in their absolute flux, and is found to be close to unity within 2\%; for the low state spectra, the constant is fixed at unity due to low statistics. The TBabs model \citep{Wilms2000} is used to model the interstellar absorption, with a Milky Way component fixed at $N_{\rm H} = 2.06 \times 10^{20}$~cm$^{-2}$ \citep{Kalberla2005} and a free, extragalactic component.  

\begin{figure}
\centering
\includegraphics[width=\columnwidth]{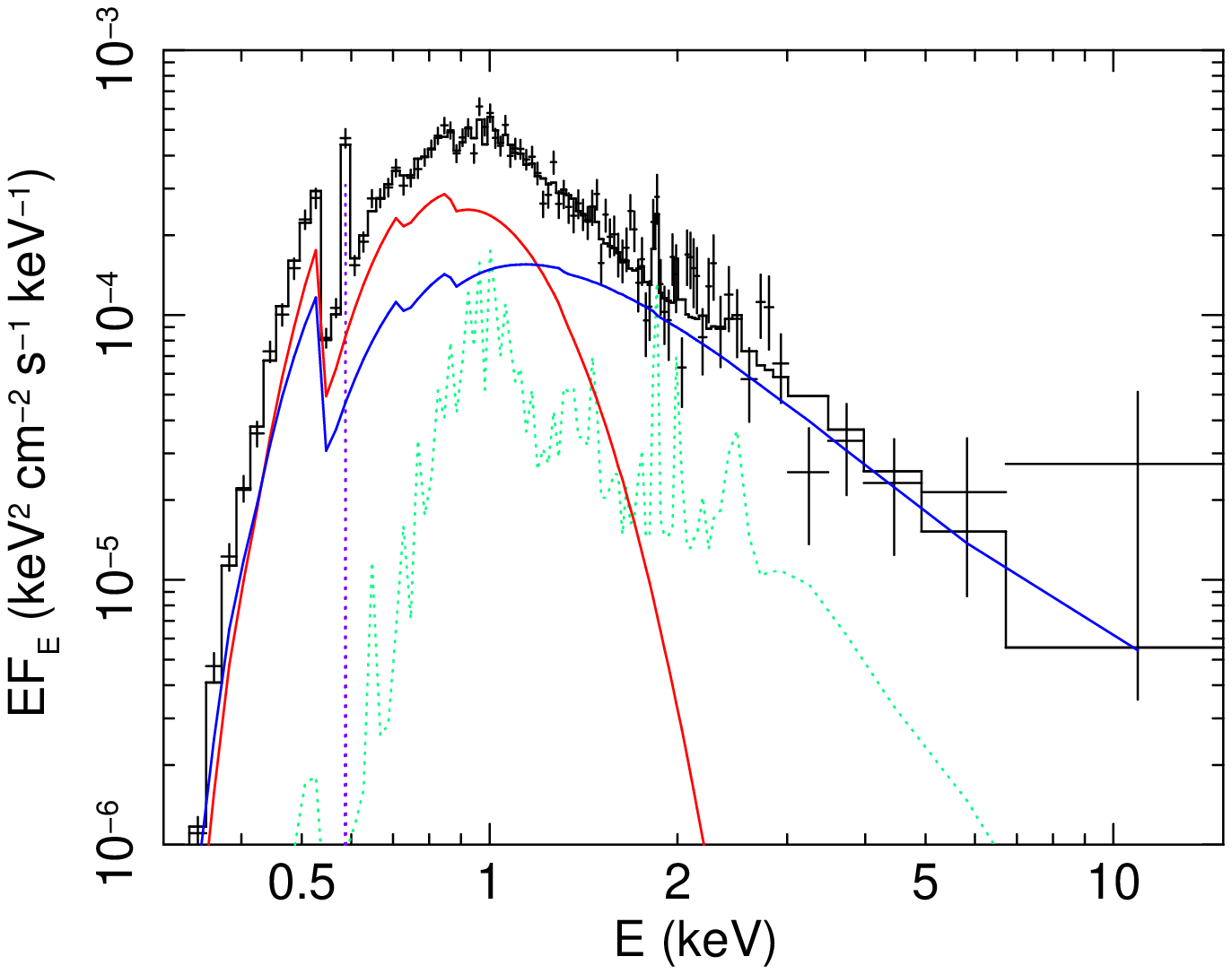}\\
\includegraphics[width=\columnwidth]{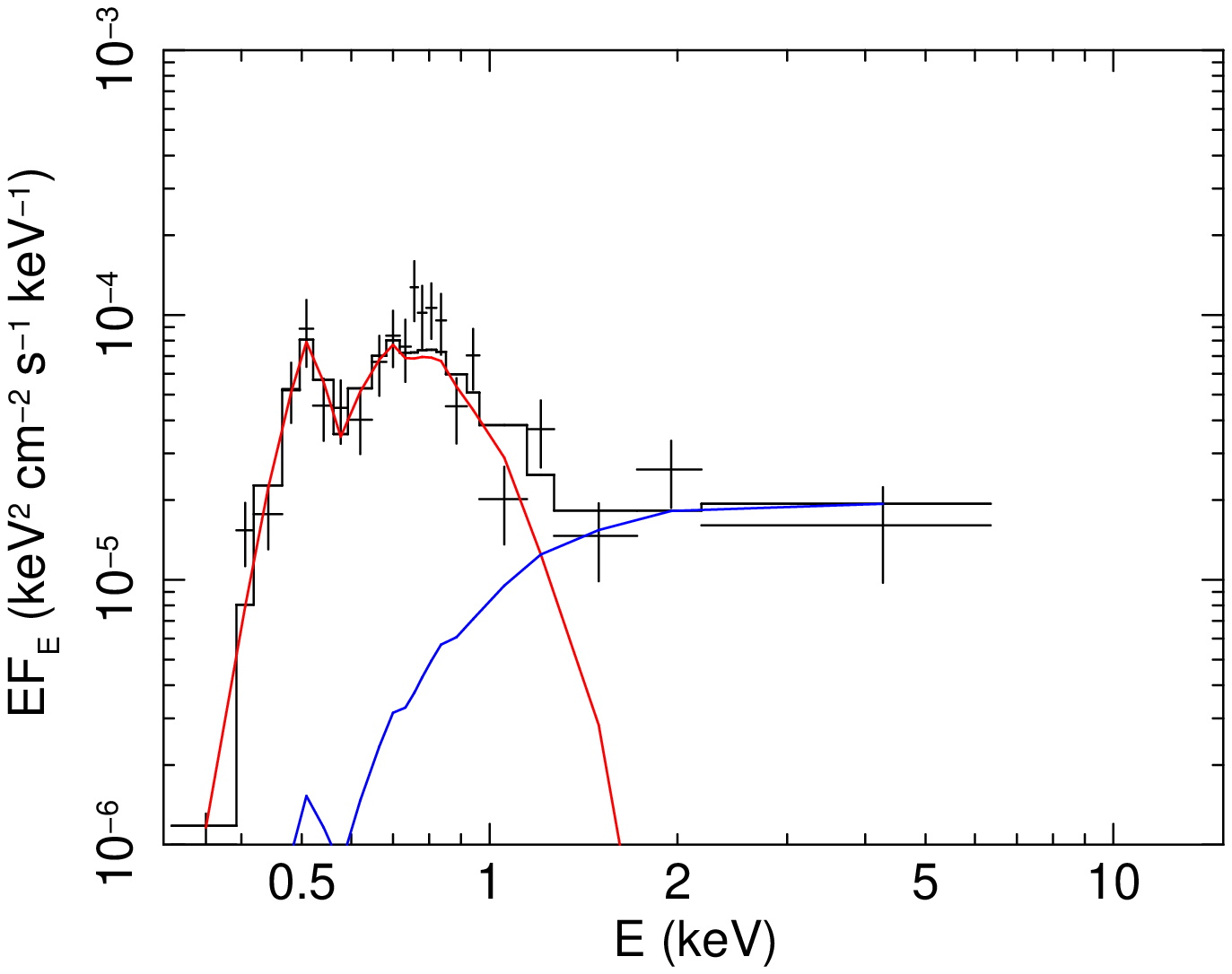}\\
\includegraphics[width=\columnwidth]{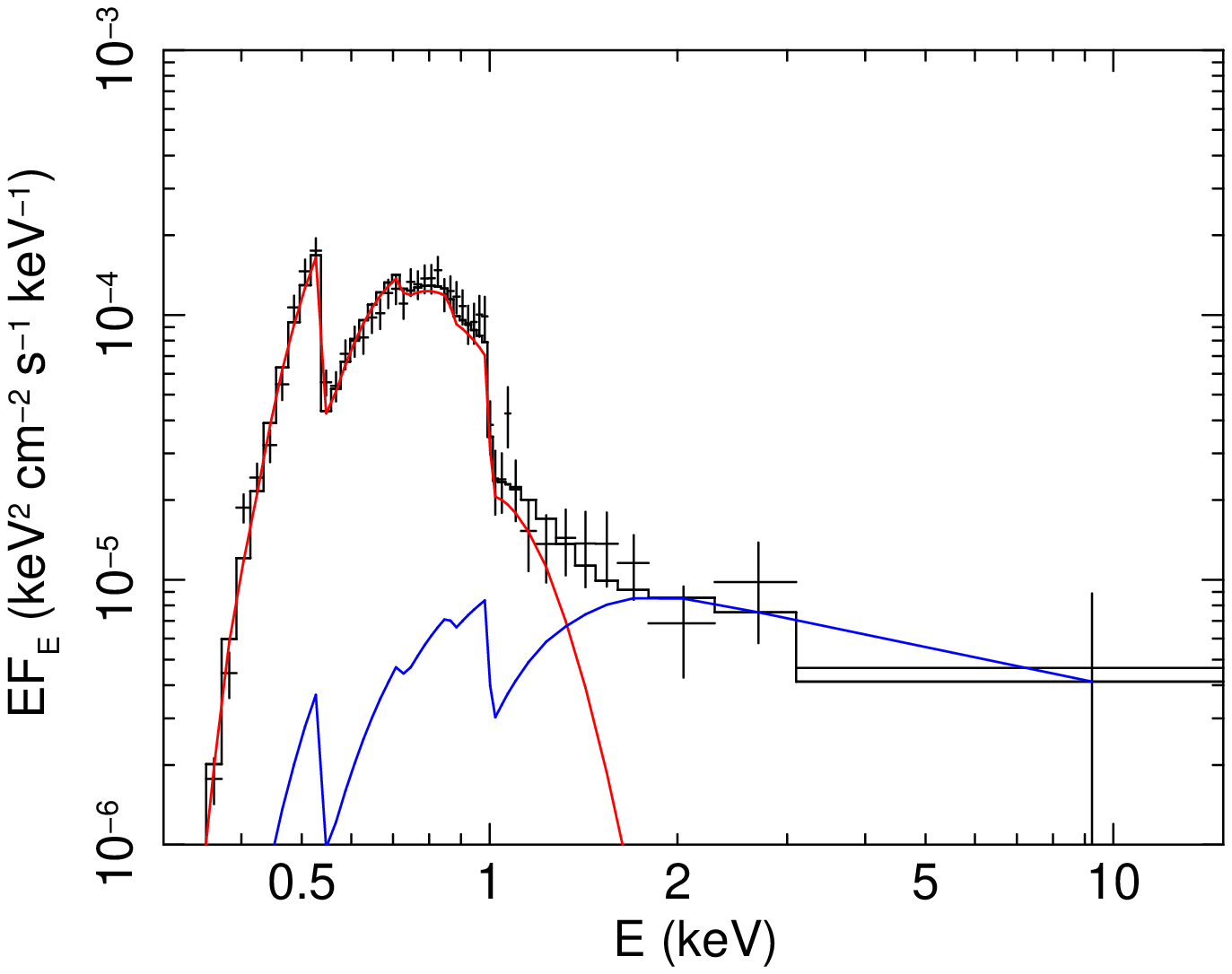}
\caption{Unfolded X-ray spectra and best-fit models for NGC 247. \textbf{Top}: the 2014 observation in the high state; the model consists of a blackbody component (red), a power-law component (blue), a thermal plasma (APEC) component (green), and an emission line (purple). \textbf{Middle}: the 2014 observation in the low state, fit with a blackbody plus power-law model with fixed absorption column density. \textbf{Bottom}: the 2009 spectrum decomposed into a blackbody plus a power-law component, both subject to an absorption edge near 1 keV. 
\label{fig:xspec}}
\end{figure}

\tabletypesize{\scriptsize}
\begin{deluxetable*}{lllllllllllll}
\tablecolumns{13}
\tabletypesize{\scriptsize}
\tablewidth{\linewidth}
\tablecaption{Spectral modeling to the {\it XMM-Newton} spectra.
\label{tab:fit}}
\tablehead{
\colhead{Model} & \colhead{$N_{\rm H,ext}^{a}$} & 
\colhead{$kT_{\rm th}$}  & \colhead{$R_{\rm th}^b$} & 
\colhead{$\Gamma$} & \colhead{$N_{\rm  PL}^c$}  &   
\colhead{$kT_{\rm APEC}$} &  
\colhead{$E_{\rm line}$} & \colhead{$N_{\rm line}^d$} &  
\colhead{$f_{\rm 0.3-10\;keV}$} & \colhead{$L_{\rm 0.3-10\;keV}$} & \colhead{$L_{\rm th, bol}$} &
\colhead{$\chi^2/{\rm dof}$} \\ 
\colhead{} & \colhead{($10^{21}$~cm$^{-2}$)} & \colhead{(keV)} & \colhead{($10^4$ km)} & 
\colhead{} & \colhead{} & \colhead{(keV)} & \colhead{(keV)} & \colhead{} & \colhead{($10^{-13}$ cgs)} & \colhead{($10^{40}$ cgs)} & \colhead{($10^{40}$ cgs)}  &\colhead{}
}
\startdata

\multicolumn{13}{c}{ObsID 0728190101: high State}\\ 
\noalign{\smallskip}\hline\noalign{\smallskip}

A & $4.5_{-0.5}^{+0.5}$ & $0.145_{-0.012}^{+0.013}$ & $1.5_{-0.4}^{+0.7}$ & 
$3.9_{-0.3}^{+0.3}$ & $4.9_{-1.3}^{+1.6}$ & \nodata & \nodata & \nodata & 
$7.90_{-0.17}^{+0.13}$ & $1.7_{-0.5}^{+0.7}$ & 1.2
& 320.02/270 \\

B & $3.7_{-0.6}^{+0.6}$ & $0.156_{-0.017}^{+0.019}$ & $0.9_{-0.4}^{+0.6}$ & 
$3.7_{-0.4}^{+0.4}$ & $3.2_{-1.1}^{+1.4}$ & $1.01_{-0.08}^{+0.12}$
 & \nodata & \nodata & $7.94_{-0.18}^{+0.07}$ & $1.0_{-0.3}^{+0.6}$ & 
0.69 & 290.40/268 \\

C & $4.2_{-0.7}^{+0.8}$ & $0.148_{-0.017}^{+0.021}$ & $1.2_{-0.5}^{+0.5}$ & 
$3.9_{-0.4}^{+0.4}$ & $3.7_{-1.3}^{+1.7}$ & $0.98_{-0.05}^{+0.12}$ & 
$0.585_{-0.011}^{+0.010}$ & $6_{-3}^{+6}$ & $7.95_{-0.21}^{+0.11}$ & 
$1.4_{-0.6}^{+1.1}$ & 0.94 & 267.48/266 \\

\noalign{\smallskip}\hline\noalign{\smallskip}

A$^\prime$ & $4.0_{-0.5}^{+0.5}$ & $0.128_{-0.009}^{+0.010}$ & 
$1.5_{-0.4}^{+0.6}$ & $3.9_{-0.3}^{+0.3}$ & $4.9_{-1.2}^{+1.5}$
 & \nodata & \nodata & \nodata & $7.91_{-0.17}^{+0.13}$ & $1.1_{-0.3}^{+0.5}$ & 
0.77 & 313.55/270 \\

B$^\prime$ & $3.5_{-0.6}^{+0.6}$ & $0.131_{-0.012}^{+0.013}$ & 
$1.2_{-0.4}^{+0.6}$ & $3.8_{-0.4}^{+0.4}$ & $3.7_{-1.1}^{+1.4}$ & 
$1.05_{-0.09}^{+0.13}$ & \nodata & \nodata & $7.93_{-0.15}^{+0.16}$ & 
$0.8_{-0.3}^{+0.5}$ & 0.56 & 290.40/268 \\

C$^\prime$ & $3.9_{-0.7}^{+0.8}$ & $0.126_{-0.013}^{+0.014}$ & 
$1.5_{-0.5}^{+1.0}$ & $4.0_{-0.4}^{+0.4}$ & $4.2_{-1.1}^{+1.2}$ & 
$1.01_{-0.09}^{+0.13}$ & $0.583_{-0.013}^{+0.013}$ & $5_{-3}^{+5}$ & 
$7.91_{-0.15}^{+0.15}$ & $1.1_{-0.4}^{+0.8}$ & 0.69
& 264.01/266 \\

\cutinhead{ObsID 0728190101: low State}

A & 4.0 fixed & $0.110_{-0.007}^{+0.007}$ & $1.7_{-0.4}^{+0.5}$ & $2.1_{-0.8}^{+0.9}$ & 
$0.22_{-0.12}^{+0.19}$ & \nodata & \nodata & \nodata & $1.6_{-0.3}^{+0.4}$ & 
$0.36_{-0.05}^{+0.05}$ & 0.56 & 42.68/29 \\

\noalign{\smallskip}\hline\noalign{\smallskip}

A$^\prime$ & 4.0 fixed & $0.092_{-0.005}^{+0.005}$ & $2.4_{-0.4}^{+0.6}$ & $2.1_{-0.7}^{+0.9}$
 & $0.25_{-0.13}^{+0.19}$ & \nodata & \nodata & \nodata & $1.5_{-0.3}^{+0.4}$ & 
$0.32_{-0.04}^{+0.05}$ & 0.54 & 39.69/29 \\

\cutinhead{ObsID 0601010101}
&&&&&&& \makebox[4em][c]{$E_{\rm edge}$} & \makebox[3em][c]{$\tau_{\rm edge}$} &&&&\\
\noalign{\smallskip}
&&&&&&& \makebox[4em][c]{(keV)} &&&&&\\
\noalign{\smallskip}\hline\noalign{\smallskip}

D & $4.5_{-0.9}^{+1.1}$ & $0.108_{-0.014}^{+0.016}$ & $2.8_{-1.5}^{+3.8}$ & 
$2.8_{-1.0}^{+1.7}$ & $0.21_{-0.12}^{+0.44}$ & \nodata & 
$0.994_{-0.018}^{+0.019}$ & $1.3_{-0.4}^{+0.4}$ & $1.68_{-0.11}^{+0.16}$ & 
$0.8_{-0.4}^{+1.2}$ & 1.4 & 129.01/110 \\

D$^\prime$ & $4.0_{-0.9}^{+1.2}$ & $0.094_{-0.012}^{+0.013}$ & 
$2.9_{-1.5}^{+3.7}$ & $3.1_{-1.1}^{+2.8}$ & $0.25_{-0.15}^{+1.15}$ & \nodata & 
$0.996_{-0.020}^{+0.022}$ & $1.2_{-0.4}^{+0.6}$ & $1.65_{-0.12}^{+0.16}$ & 
$0.5_{-0.3}^{+0.8}$ & 0.88 & 132.48/110 \\

\enddata
\tablecomments{Model A: MCD + power-law; model B: MCD + power-law + APEC; model C: MCD + power-law + APEC + Gaussian; model D: edge$\ast$(MCD + power-law).  Models A$^\prime$B$^\prime$C$^\prime$D$^\prime$: using blackbody instead of MCD.  $T_{\rm th}$ and $R_{\rm th}$ are the temperature and radius for the thermal component (MCD or blackbody), respectively; they refer to the values at the inner radius for the MCD model. $f_{\rm 0.3-10\;keV}$ is the observed flux in 0.3--10 keV; $L_{\rm 0.3-10\;keV}$ is the absorption-corrected luminosity in 0.3-10 keV; $L_{\rm th, bol} \equiv 4\pi\sigma R_{\rm th}^2 T_{\rm th}^4$ is the bolometric luminosity for the thermal component. All errors are quoted in 90\% confidence level. }

\tablenotetext{a}{Extragalactic absorption column density; an additional absorption component fixed at $N_{\rm H} = 2.06 \times 10^{20}$~cm$^{-2}$ to account for Galactic absorption is not listed here.}
\tablenotetext{b}{Assuming a face-on disk.}
\tablenotetext{c}{Normalization for the power-law component in units of $10^{-4}$~photons~keV$^{-1}$~cm$^{-2}$~s$^{-1}$ at 1~keV.}
\tablenotetext{d}{Normalization for the Gaussian component in units of $10^{-4}$~photons~cm$^{-2}$~s$^{-1}$.}

\end{deluxetable*}

\begin{deluxetable*}{lllllllll}
\tablecolumns{9}
\tabletypesize{\scriptsize}
\tablewidth{0pc}
\tablecaption{Spectral modeling to the {\it Chandra} spectra.
\label{tab:chafit}}
\tablehead{
\colhead{ObsID} & \colhead{Model} & \colhead{$N_{\rm H,ext}$} & 
\colhead{$kT_{\rm th}$}  & \colhead{$R_{\rm th}$} & 
\colhead{$f_{\rm 0.3-10\;keV}$} & \colhead{$L_{\rm 0.3-10\;keV}$} & \colhead{$L_{\rm th, bol}$} &
\colhead{$\chi^2/{\rm dof}$} \\ 
\colhead{} & \colhead{} & \colhead{($10^{21}$~cm$^{-2}$)} & \colhead{(keV)} & \colhead{($10^4$ km)} & 
\colhead{($10^{-13}$ cgs)} & \colhead{($10^{40}$ cgs)} & \colhead{($10^{40}$ cgs)}  &\colhead{}
}
\startdata

12437 & MCD & 4.0 fixed & $0.105_{-0.009}^{+0.011}$ & $3.1_{-1.0}^{+1.4}$ & $1.9_{-0.2}^{+0.2}$ & $0.8_{-0.2}^{+0.2}$ & 1.5 & 4.39/6 \\
17547 & MCD & 4.0 fixed & $0.125_{-0.011}^{+0.013}$ & $1.6_{-0.6}^{+0.8}$ & $2.0_{-0.3}^{+0.3}$ & $0.60_{-0.17}^{+0.20}$ & 0.82 & 2.10/7 \\

\noalign{\smallskip}\hline\noalign{\smallskip}

12437 & blackbody & 4.0 fixed & $0.089_{-0.007}^{+0.008}$ & $4.2_{-1.2}^{+1.6}$ & $1.9_{-0.2}^{+0.2}$ & $0.73_{-0.18}^{+0.20}$ & 1.4 & 4.11/6 \\
17547 & blackbody & 4.0 fixed & $0.105_{-0.009}^{+0.010}$ & $2.2_{-0.7}^{+0.9}$ & $1.9_{-0.3}^{+0.3}$ & $0.48_{-0.14}^{+0.17}$ & 0.76 & 3.10/7 

\enddata
\tablecomments{Parameters mean the same as in Table~\ref{tab:fit}. }
\end{deluxetable*}

\begin{deluxetable*}{llllllllll}
\tablecolumns{10}
\tabletypesize{\scriptsize}
\tablewidth{0pc}
\tablecaption{Spectral modeling to the {\it Swift}/XRT spectra.
\label{tab:xrtfit}}
\tablehead{
\colhead{group} & \colhead{ObsID} & \colhead{Model} & \colhead{$N_{\rm H,ext}$} & 
\colhead{$kT_{\rm th}$}  & \colhead{$R_{\rm th}$} & 
\colhead{$f_{\rm 0.3-10\;keV}$} & \colhead{$L_{\rm 0.3-10\;keV}$} & \colhead{$L_{\rm th, bol}$} &
\colhead{$\chi^2/{\rm dof}$} \\ 
\colhead{} & \colhead{(000334690+)} & \colhead{} & \colhead{($10^{21}$~cm$^{-2}$)} & \colhead{(keV)} & \colhead{($10^4$ km)} & 
\colhead{($10^{-13}$ cgs)} & \colhead{($10^{40}$ cgs)} & \colhead{($10^{40}$ cgs)}  &\colhead{}
}
\startdata

1 & 01/02/06 & MCD & 4.0 fixed & $0.160_{-0.015}^{+0.017}$ & $0.9_{-0.3}^{+0.4}$ & $4.0_{-0.6}^{+0.6}$ & $0.71_{-0.17}^{+0.19}$ & 0.73 & 1.50/6 \\
2 & 03/04/10/26 & MCD & 4.0 fixed & $0.139_{-0.013}^{+0.014}$ & $1.4_{-0.5}^{+0.6}$ & $3.4_{-0.5}^{+0.5}$ & $0.8_{-0.2}^{+0.2}$ & 0.96 & 4.36/6 \\
3 & 05/15/16/20/22 & MCD & 4.0 fixed & $0.17_{-0.02}^{+0.02}$ & $0.6_{-0.2}^{+0.3}$ & $2.5_{-0.4}^{+0.4}$ & $0.39_{-0.11}^{+0.13}$ & 0.38 & 5.87/5 \\
4 & 07/08/18/21/24/25/28 & MCD & 4.0 fixed &  $0.110_{-0.009}^{+0.010}$ & $2.8_{-0.8}^{+1.1}$ & $2.3_{-0.4}^{+0.4}$ & $0.9_{-0.2}^{+0.2}$ & 1.50 & 3.58/5 \\
5 & 09/11/12/13/14/17/19/23/27 & MCD & 4.0 fixed & $0.142_{-0.012}^{+0.014}$ & $0.9_{-0.3}^{+0.3}$ & $1.5_{-0.2}^{+0.2}$ & $0.34_{-0.08}^{+0.09}$ & 0.40 & 4.05/6 \\
\noalign{\smallskip}\hline\noalign{\smallskip}
1 & 01/02/06 & blackbody & 4.0 fixed & $0.130_{-0.011}^{+0.012}$ & $1.5_{-0.4}^{+0.6}$ & $4.0_{-0.6}^{+0.6}$ & $0.59_{-0.15}^{+0.17}$ & 0.79 & 2.30/6 \\
2 & 03/04/10/26 & blackbody & 4.0 fixed & $0.114_{-0.010}^{+0.011}$ & $2.1_{-0.7}^{+0.9}$ & $3.3_{-0.5}^{+0.5}$ & $0.68_{-0.18}^{+0.21}$ & 0.99 & 5.06/6 \\
3 & 05/15/16/20/22 & blackbody & 4.0 fixed & $0.138_{-0.014}^{+0.017}$ & $0.9_{-0.3}^{+0.4}$ & $2.5_{-0.4}^{+0.4}$ & $0.32_{-0.09}^{+0.11}$ & 0.41 & 7.83/5 \\
4 & 07/08/18/21/24/25/28 & blackbody & 4.0 fixed & $0.094_{-0.007}^{+0.007}$ & $3.8_{-1.0}^{+1.3}$ & $2.3_{-0.4}^{+0.4}$ & $0.78_{-0.19}^{+0.21}$ & 1.40 & 3.01/5  \\
5 & 09/11/12/13/14/17/19/23/27 & blackbody & 4.0 fixed & $0.117_{-0.009}^{+0.010}$ & $1.3_{-0.4}^{+0.5}$ & $1.5_{-0.2}^{+0.2}$ & $0.29_{-0.07}^{+0.08}$ & 0.42 & 3.46/6 
\enddata
\tablecomments{Parameters mean the same as in Table~\ref{tab:fit}. }
\end{deluxetable*}

Following \citet{Jin2011}, two models were used to fit the soft thermal emission, a multicolor disk \citep[MCD;][]{Shakura1973} and a mono-temperature blackbody. A power-law component is added to account for the hard tail in the spectrum. A combination of a MCD and a power-law is standard for modeling the X-ray spectrum of accreting black hole binaries, while the choice of a mono-temperature blackbody is appropriate if the soft thermal emission arises from the photosphere of thick outflows. 

For the high state spectra from the 2014 observation (ObsID 0728190101), the fitting is significantly improved if an optically-thin thermal plasma model (APEC) is added to the two-component model. The F-test suggests that the inclusion of the APEC model has a chance probability of $\sim3 \times 10^{-6}$. Inspection of the residuals, especially for the PN data, shows an excess around 0.6 keV. To account for that feature we added a zero-width Gaussian; the F-test chance probability is $2 \times 10^{-5}$. 

For the low state spectra in the 2014 observation, the MCD/blackbody plus power-law model gives an adequate fit and presence of additional components cannot be justified due to the limited statistics. The low statistics also prevent the fitting from breaking the degeneracy between the absorption column density and the temperature of the thermal component. Keeping both parameters free, the fitting converges to a combination of strong absorption ($\sim 10^{22}$~cm$^{-2}$) and a cool, large-area soft thermal component with an unphysically large bolometric luminosity ($\sim 10^{42}$~\ergs).  Thus, we fixed the extragalactic absorption column density at $4 \times 10^{21}$~cm$^{-2}$, the typical value derived from the high-quality data. If we adopt the best-fit model derived from the high-state spectra and fix all parameters except the extra-galactic absorption column density, no successful fit can be obtained, with a reduced $\chi^2$ $\sim$ 13. Thus, we conclude that the dipping is not a result of increased absorption alone.

The spectral analysis for the 2009 observation (ObsID 0601010101) has been reported in detail in \citet{Jin2011}. The best-fit spectral model consists of two major components, a blackbody or MCD component dominant below 2 keV and a power-law component dominant above 2 keV. In addition, inclusion of an absorption edge near 1 keV improved the fit, with a chance probability of $1 \times 10^{-4}$ estimated via simulation. Thus, for consistency, we extracted the energy spectra in the same manner as above and fitted them using the best-fit model found by \citet{Jin2011}, i.e., edge$\ast$(MCD/blackbody + power-law) subject to both Galactic and extragalactic absorption (TBabs). Here we use both PN and MOS data while \citet{Jin2011} used PN data only. Our results are consistent within 1 or 2$\sigma$ errors. The X-ray spectra with model components are displayed in Figure~\ref{fig:xspec} and all the model parameters are tabulated in Table~\ref{tab:fit}. The observed flux and intrinsic luminosity was calculated using the {\tt cflux} command in XSPEC. The factor for absorption-correction is $\sim$10 for the 2014 high state spectra, $\sim$15 for the low state spectra, $\sim$20 for the 2009 spectra.

For the 2014 high state spectra, 45\% of the intrinsic luminosity in the energy range of 0.3--10 keV arises from the blackbody component, 45\% from the power-law component, 3\% from the APEC component, and 6\% from the Gaussian line.  For the 2014 low state and 2009 spectra, the blackbody component is dominant and has $\sim$95\% of the luminosity in the same band. 

For the {\it Chandra} spectra, a single blackbody or MCD spectrum subject to interstellar absorption (TBabs) is adequate to fit the spectrum in the 0.3--1 keV band (for ObsID 12437) or the 0.3--2 keV band (for ObsID 17547).  The best-fit results are listed in Table~\ref{tab:chafit}.  The extragalactic absorption column density is fixed at the typical value, $4 \times 10^{21}$~cm$^{-2}$, found from the {\it XMM-Newton} data. If it is allowed to vary, the best-fit value is around $(2-5) \times 10^{21}$~cm$^{-2}$ but the error bounds of this and other parameters cannot be well constrained, due to the limitation of data quality.  Each combined {\it Swift}/XRT spectrum is fitted in the same way as for the {\it Chandra} spectra. The results are presented in Table~\ref{tab:xrtfit}.

\subsection{Long-term X-ray variability and spectral evolution}

\begin{figure}
\centering
\includegraphics[width=\columnwidth]{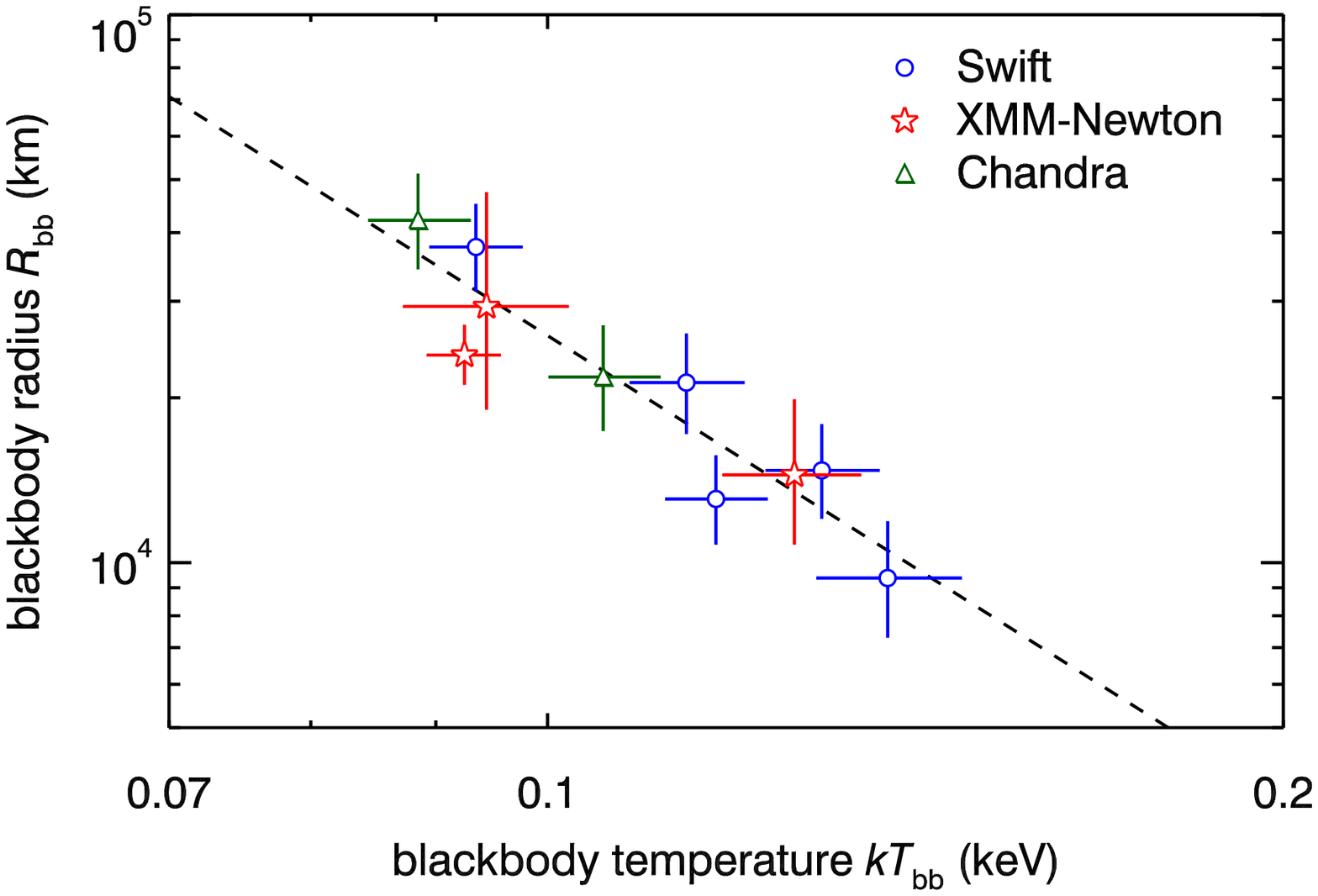}\\
\includegraphics[width=\columnwidth]{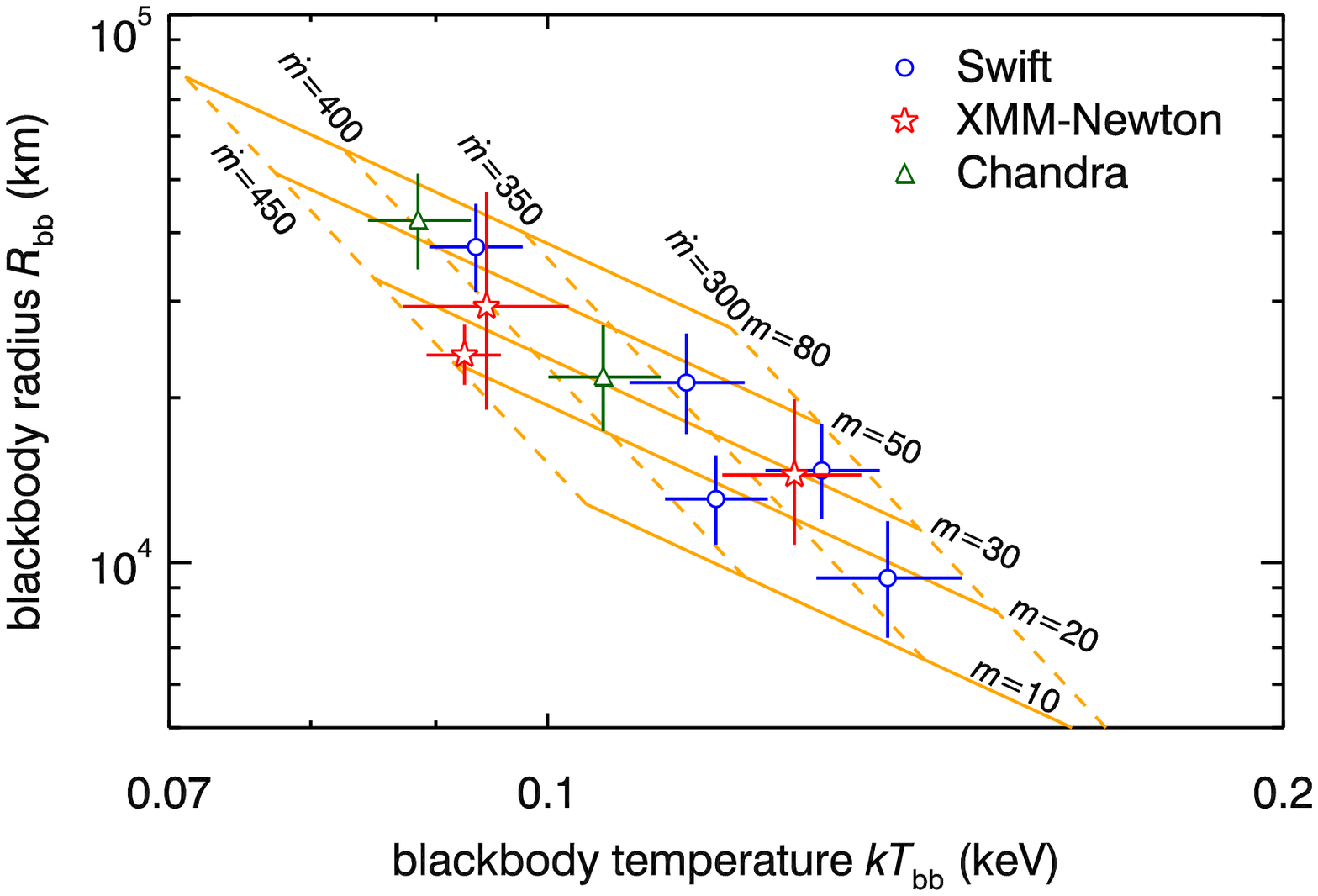}
\caption{Blackbody radius versus temperature for NGC 247 ULX. {\bf Top}: data and linear regression (dashed line) with a slope of $-2.8 \pm 0.3$. {\bf Bottom}: data and predicted radius-temperature relations from an outflow model \citep{Soria2016} assuming depleted hydrogen and half of the accretion mass channeled to wind. $m$ is the black hole mass in units of $M_\sun$, and $\dot{m}$ is the mass accretion rate in units of the Eddington rate. Solid lines represent $R-T$ relations given a black hole mass of 10, 20, 30, 50, or 80 $M_\sun$, and each dashed line connects black holes of different masses at the same accretion rate, at 300, 350, 400, or 450 times the Eddington rate. 
\label{fig:rt}}
\end{figure}

The best-fit blackbody temperature and radius for each observation or observation group is shown in Figure~\ref{fig:rt}.  For {\it XMM-Newton} data, results from the most sophisticated models (the best model) are adopted and the absorption column density is a free parameter, while for {\it Chandra} and {\it Swift} data, due to the low quality of data, the column density is fixed.  The small errors on {\it Chandra} and {\it Swift} results are due to the fact that the absorption component was fixed and its large degeneracy with the soft blackbody component was vanished. A linear correlation is seen between the blackbody temperature and radius on logarithmic scale. A linear regression using the bivariate correlated errors and intrinsic scatter (BCES) algorithm \citep{Akritas1996}, which considers errors on both $x$ and $y$ in the type of $y|x$ (to predict $y$ using $x$), suggests a relation $R_{\rm bb} \propto T_{\rm bb}^{-2.8 \pm 0.3}$.  We also plot model predicted relations assuming super-Eddington accretion with massive outflows \citep{Soria2016} to compare with the data.  Assuming a hydrogen mass fraction of 0 \citep[cf.\ the supersoft ULX in M101 whose companion is likely a hydrogen depleted Wolf-Rayet star;][]{Liu2013} and 1/2 of accretion mass going to a wind, a 30 $M_\sun$ black hole accreting at 300--450 times of the Eddington rate matches the data. With a hydrogen mass fraction of 0.73 (solar abundance), the best match suggests a black hole mass of $\sim$50 $M_\sun$ at similar accretion rates (not shown in the plot). If a smaller fraction of mass is transferred to the wind, a higher accretion rate is required.  These estimates may be quantitatively inaccurate, because of the simplification of the model and a deflected $R-T$ slope. Qualitatively, a picture involving a stellar mass black hole with supercritical accretion matches the observations.

\begin{figure}
\centering
\includegraphics[width=\columnwidth]{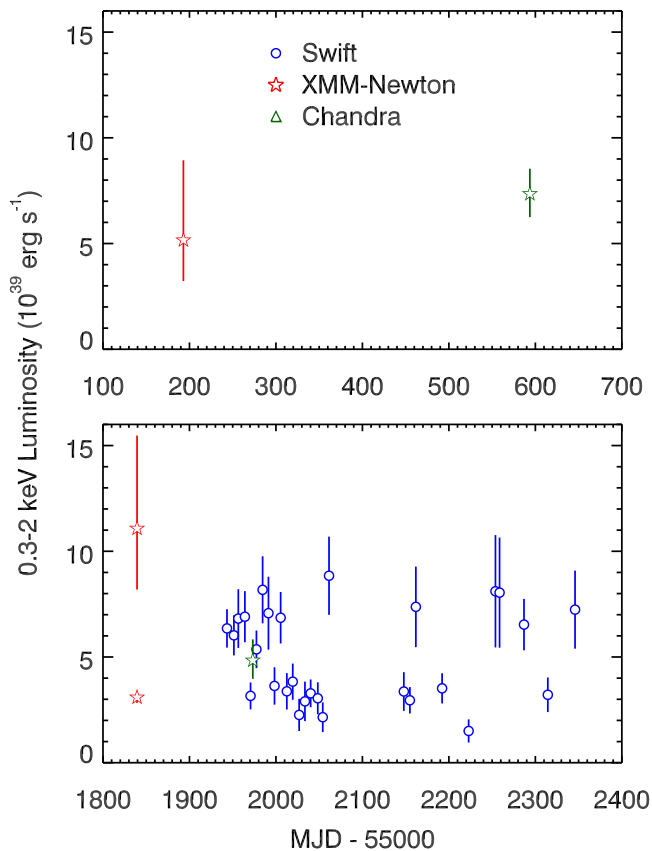}
\caption{Long-term X-ray lightcurve of NGC 247 ULX in the energy range of 0.3--2 keV. 
\label{fig:lcall}}
\end{figure}

The source lightcurve for intrinsic luminosity in the energy range of 0.3--2 keV is shown in Figure~\ref{fig:lcall}. We note that for each individual {\it Swift} observation, the luminosity is converted from the net count rate based on the best-fit model in the associated observation group (Table~\ref{tab:xrtfit}). The fractional Poisson error on the net count rate is adopted for {\it Swift} data and is certainly an underestimate. For {\it XMM-Newton} and {\it Chandra} data, the luminosities and errors are directly computed from their best-fit models using the {\tt cflux} model in XSPEC.  The source is not as variable as other soft ULXs like the one in M101. No specific pattern or period can be found in the current lightcurve. 

\subsection{Short-term X-ray variability}

\begin{figure}
\centering
\includegraphics[width=\columnwidth]{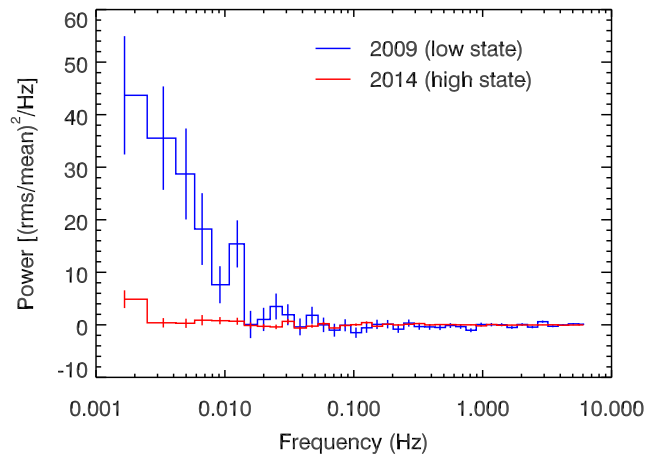}
\caption{X-ray power spectrum density in the 0.3-2 keV band from the 2009 {\it XMM-Newton} observation (blue), where the flux is relatively low, and from the 2014 observation during the non-dipping state (red), where the flux is relatively high. 
\label{fig:psd}}
\end{figure}

The Fourier power spectrum in the energy band of 0.3--2 keV for the 2009 {\it XMM-Newton} observation was reported in \citet{Jin2011}.  Here we include timing analysis of the 2014 observation. The power spectrum was generated from time intervals free of background flares and timing gaps, using PN events in the source region with FLAG = \#XMMEA\_EP and PATTERN $\le$ 4. The lightcurves were generated with a resolution of the CCD frame time ($\sim$73 ms) and divided into short segments for 8192-point Fourier transforms.  The final power spectrum density was obtained by coadding spectra from individual segments, normalized to root-mean-square (rms) with Poisson noise subtracted, and binned in logarithmic scale on frequency.

The power spectra for the 2009 observation and for the 2014 observation during the high (non-dipping) state are displayed in Figure~\ref{fig:psd} for comparison. For the dipping intervals, there are not sufficient counts for timing analysis.  It is obvious that the source displayed huge short-term variability in the 2009 observation (low state) but much less in 2014 (high state). The fractional rms variability is $0.49 \pm 0.03$ in 2009 (low state) and $0.12 \pm 0.02$ in 2014 (high state), both in the frequency range of 1.66~mHz to 0.01 Hz and in the energy range of 0.3-2 keV. The variability in the 2009 observation was energy dependent \citep{Jin2011} with a fractional rms of $0.33 \pm 0.04$ in the 0.3--0.8 keV band and $0.76 \pm 0.07$ in the 0.8--2 keV band. For the high state in the 2014 observation, the variability seems to be energy independent, with a fractional rms of $0.15 \pm 0.03$ in the 0.3--0.8 keV band and $0.11 \pm 0.03$ in the 0.8--2 keV band. At energies above 2 keV, no significant power or meaningful upper limit can be obtained due to the low statistics for both observations. 


\subsection{Spectral modeling to the optical data}

\begin{figure}
\centering
\includegraphics[width=\columnwidth]{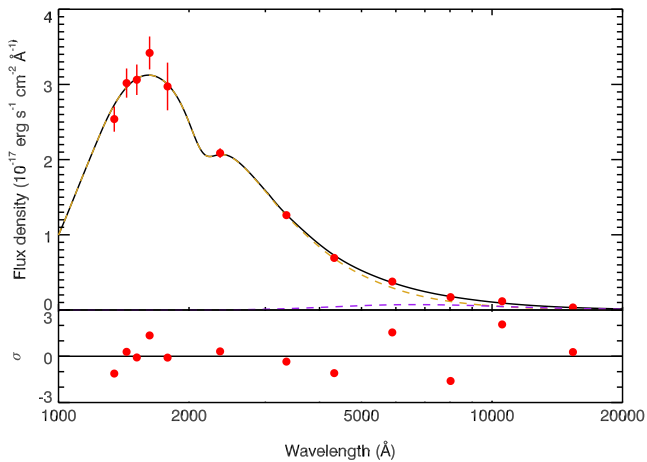}
\caption{Measured {\it HST} spectrum for the ULX using WFC3 photometry and ACS/SBC grism data.  The model (solid black) consists of two blackbody components (dashed yellow and purple), see Table~\ref{tab:hst}.
\label{fig:opt}}
\end{figure}

The measured {\it HST} spectrum composed of the WFC3 and SBC data is shown in Figure~\ref{fig:opt}. A combination of two blackbody components subject to reddening \citep{Cardelli1989} can fit the spectrum, resulting in $\chi^2$ of 91.7 with 7 degrees of freedom. We note that most of the residuals come from the two IR data points.  Excluding the IR data gives $\chi^2$ of 6.7 with 5 degrees of freedom for the same model.  Of course the fit can be improved by adding more spectral components, but that is hard to justify based on two data points. As we mentioned above, the IR flux may have extra uncertainties due to systematic errors in photometry. If we assume a systematic error of 10\% in flux, the fitting is acceptable with $\chi^2$ of 13.9 and 7 degrees of freedom. We adopt this as the formal results and the best-fit model parameters are listed in Table~\ref{tab:hst}. For comparison, a reddened power-law model gives a poor fit with a $\chi^2$ of 547 with 10 degrees of freedom. 

\tabletypesize{\footnotesize}
\begin{deluxetable}{ll}
\tablecolumns{2}
\tablewidth{\linewidth}
\tablecaption{Model parameters fitted to the {\it HST} spectra.
\label{tab:hst}}
\tablehead{
\colhead{Parameter} & \colhead{Value}
}
\startdata
Extinction $E(B-V)$ & $0.085 \pm 0.017$\\
blackbody temperature (K) & $19000 \pm 400$ \\
blackbody area (cm$^2$) & $(2.5 \pm 1.7) \times 10^{25}$ \\
blackbody temperature (K) & $4600 \pm 400$ \\
blackbody area (cm$^2$) & $(4.5 \pm 1.6) \times 10^{26}$ \\
$\chi^2$/dof & 13.9/7 
\enddata
\tablecomments{Assuming a 10\% of uncertainty on the WFC3/IR flux. An assumption of 20\% will change the $\chi^2$ to 9.4 but do not affect the derived error bounds, which are quoted at 1$\sigma$ confidence level.}
\end{deluxetable}

\begin{figure}
\centering
\includegraphics[width=\columnwidth]{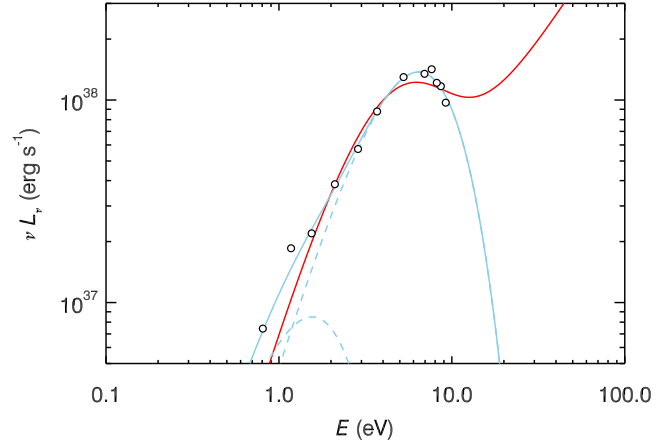}
\caption{Fitting the optical data with an self-irradiated standard accretion disk model (red). The circles are unreddened optical data and blue lines are the two-component blackbody model fitted to the data (Figure~\ref{fig:opt}).  
\label{fig:imcd}}
\end{figure}

Even at extremely high accretion rates, the outer part of the accretion disk is still thought to keep the form predicted by the standard accretion disk model \citep{Shakura1973}, which predicts a power-law spectrum, $F_\nu \propto \nu^{1/3}$, in the optical band and cuts off at $\nu_{\rm out}$ corresponding to the temperature of the blackbody emission at the outermost radius $R_{\rm out}$.  If irradiation from the inner disk onto the outer disk is taken into account \citep[e.g.,][]{Gierlinski2009,Sutton2014}, it changes the radial temperature profile and also predicts a power-law like spectrum at frequencies above $\nu_{\rm out}$, except that the spectral energy distribution displays a bump before rolling down at $\nu_{\rm out}$.  To test if the observed blackbody-like optical spectrum is a result of disk irradiation, we fit the optical data below 10000~\AA\ with a self irradiated standard disk model \citep{Madhusudhan2008} with three parameters: the X-ray (irradiating) luminosity, and the inner and outer disk radius. We will show later in the discussion section that the MCD model used to fit the soft thermal spectrum is invalid. If the inner accretion disk is covered by optically-thick outflows, the disk inner radius cannot be directly measured from the spectrum. We thus fix the inner radius at $10^7$~cm typically for a 10 $M_\sun$ stellar mass black hole. The best-fit results are shown in Figure~\ref{fig:imcd}, with an irradiating luminosity of $\sim10^{41}$~\ergs\ and an outer radius of $\sim 10^{12}$~cm. 

If we consider the WFC3 data only, a power-law model ($F_\lambda \propto \lambda^{-2.9 \pm 0.1}$ with $E(B-V) = 0.18 \pm 0.02$) provides a better fit than a blackbody, with a $\chi^2$ of 8.2 versus 97.0 with 4 degrees of freedom. However, the reddened power-law model predicts a UV flux higher than the observed flux in the ACS SBC band by a factor of 1.7--3.5.  Inclusion of UV data below 2000\AA\ is key in determining the optical emission mechanism. 

\section{Discussion}

These multiwavelength and multi-epoch observations provide insight into the nature of the soft ULX in NGC 247. The physical interpretation of the data and results are briefly discussed in the following. 

\subsection{Transition between the supersoft and soft ultraluminous regimes}

In contrast to previous observations, soft blackbody emission does not dominate the 2014 high state spectrum -- the blackbody and power-law components exhibit almost the same luminosity in the 0.3--10 keV band. The power-law component is much stronger than it was in the spectrum of other soft ULXs and NGC 247 ULX in the low state. Also, the strong fast variability previously seen in the X-ray low state is also substantially depressed in the X-ray high state. Both the spectral and temporal properties in the 2014 high state are consistent with those defined for the soft ultraluminous (SUL) regime \citep{Sutton2013}, except that the power-law component is relatively weaker and steeper and the soft blackbody component is relatively more variable on short timescales.  The X-ray high state of NGC 247 ULX may be an extreme case of the SUL regime \citep{Sutton2013}.

The X-ray low state of NGC 247 ULX, both in spectral and timing properties, defines a distinct accretion regime. We call it the {\it supersoft ultraluminous} (SSUL) regime, with a spectrum characterized by a dominant soft blackbody component typically of 0.1~keV in temperature and $10^4 - 10^5$~km in radius. In the SSUL regime, prominent short-term variability may be associated with the soft X-rays, also detected in M101 ULX \citep{Mukai2003,Mukai2005,Soria2016,Urquhart2016}. For both sources in the SSUL regime, the variability is scaled with energies with a larger fractional rms seen in higher energies toward 2 keV (the timing behavior is unknown for photons above 2 keV), which is similar to the behavior of NGC 5408 X-1 and NGC 6946 X-1 \citep{Hernandez-Garcia2015}. The origin of the fast variability seen in the SSUL regime is unclear. Due to the large size ($10^4-10^5$~km) of the blackbody emitting region and the high amplitude ($\sim$50\%) of the time variation associated with it, it seems difficult to explain it with the same scenario applied for the SUL regime, that the turbulence in the wind causes the variability by scattering hard X-rays out of the line of sight \citep{Takeuchi2013}. We speculate that the SSUL regime has an accretion rate that is even higher than in the SUL/HUL regimes. SSUL sources may transition to SUL when the accretion rate decreases. \citet{Urquhart2016} predicted that such a transition may occur when the blackbody temperature is above $\sim$150~eV and that is indeed observed in this source. 

An alternative interpretation is that the SSUL regime is simply due to a geometric effect, being a standard broad-band ULX viewed at a high inclination angle \citep[also see discussions in][]{Urquhart2016}.  The temperature of the photosphere is uneven, perhaps hotter near the disk plane and cooler outside. An occulter with a size comparable to the photosphere, such as a warped disk or a circumbinary disk far outside, may obscure the hotter region of the photosphere, leading to a lower temperature and flux.  The motion of the occulter may be the source of the strong variability.

There is no solid evidence for or against either speculation. The link between the absorption edge and the strong variability may be explained by both, as due to an increased wind caused by increased accretion or related to obscuration. The power spectrum density in a $\nu^{-1}$ shape consistent with that produced by viscous instability, along with the negative $R-T$ relation, is in favor of the high accretion scenario.  Similar to NGC 247 ULX, NGC 5408 X-1 displays a negative rms-flux relation on long timescales of years \citep{Caballero-Garcia2013}, except that their measurements were made in the energy range of 1-10 keV. The energy dependence of the rms variability below 2 keV is also similar for the two sources.  These similarities seem to be in support of the viewing angle interpretation. More observations and studies of a large sample is needed to further constrain the nature of these supersoft ULXs.

On the contrary, variable ULXs like NGC 5408 X-1 and NGC 6946 X-1 show a positive rms-flux relation on short timescales, over which the timing and spectral properties are stationary \citep{Heil2010,Hernandez-Garcia2015}; this relation is found to be ubiquitous in bright accreting objects. A direct comparison can be made between this ULX and NGC 5408 X-1 if the short-term rms-flux relation can be measured in the future.

\subsection{Invalidity of the disk interpretation}

The thermal state of black hole binaries is characterized by a spectrum consisting of a dominant thermal MCD component plus a minor power-law or Comptonization component \citep{Remillard2006}.  It has been suggested that the thermal spectrum seen in these soft ULXs could be due to emission from a standard accretion disk surrounding an intermediate-mass black hole. For most other soft ULXs, this has already been ruled out because the pattern of their spectral evolution \citep[e.g., decreasing radius with increasing temperature;][]{Soria2016,Urquhart2016} is inconsistent with that seen in the thermal state (constant inner radius with varying temperature).  For NGC 247 ULX, previous studies could not rule this out due to insufficient data. Here, multi-epoch observations with {\it XMM-Newton}, {\it Chandra}, and {\it Swift} (Figure~\ref{fig:rt}) identify a similar spectral evolution pattern,  $R_{\rm bb} \propto T_{\rm bb}^{-2.8 \pm 0.3}$, which is steeper than that found for the ULX in M101 \citep[slope $\sim2$;][]{Soria2016} and for the whole population of soft ULXs \citep[slope = $-2.2 \pm 0.5$;][]{Urquhart2016}, but is consistent within errors. In the spectral fitting we had to fix the absorption column density for the {\it Chandra} and {\it Swift} data, otherwise the error bound for the blackbody radius cannot be well constrained. Such a practice seems to be justified based on the {\it XMM-Newton} data, but the slope would change if the true absorption varies due to the degeneracy between the blackbody component and absorption. 

\subsection{Connection between soft ULXs and soft excesses in broad-band ULXs}

Besides the soft ULXs, similar anti-correlations have been found in broad-band ULXs for their soft emission component (i.e., the soft excess) if it is modeled using an MCD model (similar results can be obtained if a blackbody model is used). For NGC 1313 X-2, \citet{Feng2007_NGC1313X2} found a relation $L_{\rm disk} \propto T_{\rm in}^{-7 \pm 3}$ corresponding to $R_{\rm in} \propto T_{\rm in}^{-5 \pm 1}$.  \citet{Feng2009} observed that the disk inner radius of NGC 5204 X-1 seems to shrink at higher disk inner temperatures, but the uncertainties did not allow a firm conclusion. \citet{Kajava2009} found that the soft component in broad-band ULXs follow a relation $L_{\rm disk} \propto T_{\rm in}^{-3.5}$ or  $R_{\rm in} \propto T_{\rm in}^{-4}$.  Apparently, the $R-T$ relation for these soft excesses is much steeper than for the soft ULXs. However, the situation for broad-band ULXs may be complicated, as a positive correlation consistent with a $L \propto T^4$ relation was reported for the soft component in some luminous ULXs including NGC 5204 X-1 \citep{Miller2013}.  The complexity in broad-band ULXs could be due to the importance of the hard emission; how it is modeled substantially affects the parameters derived for the soft component. The treatment of absorption is also crucial to the characterization of the soft component.  

Given a negative $R-T$ relation, the most plausible explanation is that these emission components may arise from the photosphere of massive outflows associated with supercritical accretion \citep{Poutanen2007}. If the $L \propto T^4$ relation is true, it is consistent with a scenario that the emission comes from cool disks around more massive black holes. One of the major differences 
between the SSUL regime the spectral/timing regimes of broad-band ULXs is that most soft excesses seen in broad-band ULXs are less variable on short or long timescales than the soft blackbody component seen in the SSUL regime. A possible explanation may be that the blackbody-like emission from these two classes of sources may both be associated with outflows due to supercritical accretion, but they happen at different accretion rates (supersoft ULXs may have a higher accretion rate). Alternatively, as mentioned above, the two emission components may be identical under the scenario of the geometric interpretation.

\subsection{Signatures of outflows}

In supercritical accretion models, massive outflows are a key feature predicted by theories and numerical simulations \citep{King2003,Poutanen2007,Ohsuga2011,Dotan2011,Kawashima2012,Jiang2014,King2016}.  Observationally, the detection of soft X-ray spectral residuals \citep{Middleton2014,Middleton2015b} and emission/absorption lines \citep{Pinto2016} in ULXs have been argued as a signature of outflows associated with supercritical accretion. An absorption edge near 1 keV is detected in the 2009 {\it XMM-Newton} observation of NGC 247 ULX, and is also detected from several other soft ULXs \citep[Antennae, M51, M101, and NGC 4631; see Table~4 in][]{Urquhart2016}. Such a feature may be common in these sources and is interpreted as a result of absorption by clumpy winds \citep{Urquhart2016}. The edge is likely due to L-edge absorption by highly ionized iron (e.g., Fe\,{\sc viii}). 

During the high state of the 2014 {\it XMM-Newton} observation of NGC 247 ULX, the source exhibited an emission component which can be explained by collisionally-ionized optically-thin thermal plasma. Similar features are also observed in other soft ULXs \citep[M51 and M101; see Table~4 in][]{Urquhart2016}.  The origin of the thermal plasma is still unknown. It may be due to shocks generated by wind-wind interaction or between the disk wind and stellar wind from the companion star.  The normalization of the APEC model ($N_{\rm APEC} \sim 10^{-4}$ in XSPEC) suggests that $\int{n_{\rm e} n_{\rm H} {\rm d}V} = 1.4 \times 10^{61}$.  The thermal plasma emitting region should lie outside of the photosphere of the outflow and have a lower density than the photosphere.  The density of the photosphere can be estimated as $\kappa_{\rm es} \rho_{\rm ph} r_{\rm ph} \sim 1$, where $\kappa_{\rm es} \approx 0.34$~g~cm$^{-2}$ is the electron scattering opacity. In the 2014 high state ($r_{\rm ph} \approx 1.5 \times 10^{9}$~cm), the density of at the photosphere can be inferred as $\rho_{\rm ph} \approx 2 \times 10^{-9}$~g~cm$^{-3}$. Taking this as an upper limit, the thermal plasma emitting region has a radius of at least $10^{10}$~cm. The energy of the line at 0.58~keV is consistent with emission from highly ionized oxygen (O\,{\sc vii} at 0.574~keV), whose cooling rate peaks at a temperature (0.17~keV) close to the detected blackbody temperature (0.13~keV).  The density in the line emission region $n_{\rm H} = \sqrt{L_{\rm line}/(5CR^3)}$, where $L_{\rm line} \approx 7 \times 10^{47}$~photons~s$^{-1}$ is the Gaussian line luminosity and $C = 2.16 \times 10^{-15}$~photons~cm$^3$~s$^{-1}$ is the cooling rate at the blackbody temperature. Following the same assumption above that the density is lower than that at the photosphere, the line emission radius is at least $3 \times 10^{10}$~cm.


The anti-correlation between the blackbody temperature and radius is predicted by outflow models. Adopting the wind model from \citet{Soria2016}, we find that a stellar mass black hole with a mass accretion rate $\sim10^2$ times the Eddington rate ($\equiv L_{\rm Edd} / 0.1c^2$) seems to be a plausible solution. The inferred mass of the black hole is consistent with that measured for IC 10 X-1 \citep{Silverman2008} and the binary black holes associated with the gravitational wave event GW150914 \citep{Abbott2016}, and can be explained as a result of core-collapse of stars of low metallicity. The accretion rate is extremely high but comparable to or lower than the estimated accretion rate for the Galactic microquasar SS~433 \citep{Shklovskii1981}.


\subsection{Nature of the flux dipping}

During the 2014 {\it XMM-Newton} observation, the source displayed dips in its lightcurve (Figure~\ref{fig:lc}). The transition between the high flux state and the low state occurs on a timescale of $\sim$200~s. X-ray flux dips have been observed in other ULXs, e.g., in broad-band ULXs in NGC 55 \citep{Stobbart2004}, NGC 5408 \citep{Grise2013}, and M94 \citep{Lin2013}, and also in soft ULXs in M51 and M81 \citep{Urquhart2016}. Their nature is not determined. Here the high quality {\it XMM-Newton} data allows us to see the hardness change during the dips.  It is clearly seen in Figure~\ref{fig:lc} that the blackbody temperature becomes low when entering the dip and then goes high when leaving the dip, in a manner smoother than the change of flux.  The X-ray spectrum and flux in the dips are similar to those in the 2009 {\it XMM-Newton} spectrum.  Both the spectral fitting and the change of hardness ratio clearly argue against the interpretation that the dips are purely due to absorption or partial absorption, further supported by the change of the blackbody temperature and emitting size (cooler and larger in the dips). Therefore, the dips are likely caused by the change of the properties of the photosphere itself. Again, a large and variable equatorial occulter could be another plausible explanation for the dips.

We estimate the physical timescales associated with dipping as follows. The wind launching radius or the spherization radius is $r_{\rm sp} \approx \dot{m} \, r_{\rm in}$. Assuming $\dot{m} \approx 10^2$ and a disk inner radius $r_{\rm in} \approx 10^7$~cm for a stellar mass black hole, the spherization radius is on the order of $10^9$~cm, close to the photosphere radius of the wind, $r_{\rm ph} \approx (1.5-3) \times 10^9$~cm. The photosphere density can be estimated assuming $\kappa_{\rm es} \rho_{\rm ph} r_{\rm ph} \sim 1$. The density increases as $r^{-2}$ and the temperature increases as $r^{-3/4}$ to $r^{-1/2}$ toward the base of the wind. If we assume a thin disk ($H/R \sim 0.1$) then it has a thickness of $10^8$~cm, self-consistent in the framework of \citet{Jiang2014}. From the base of the wind to the center of the disk, the ratio of the radiation pressure to the gas pressure drops by a factor of $10^3$ \citep{Jiang2014}. We also assume that the central disk temperature is on the order of 1 keV. This gives a viscous timescale of roughly $5 \times 10^3$~s and a thermal timescale of roughly $10^2$~s at $r \approx 10^9$~cm. As the viscous timescale $\tau_{\rm visc} \approx (1 / \alpha)\,(H/R)^2\,(H/c_{\rm s})$ is strongly dependent on the disk thickness, for a thick disk ($H/R \sim 1$) the viscous timescale and the thermal timescale could be on the same order of magnitude and both close to $10^2 - 10^3$~s. However, even an order-of-magnitude estimate would be difficult at such a high accretion rate, too little is understood about the disk structure. We note that \citet{Jiang2014} assumed an accretion rate of 22 times the Eddington rate and the simulation only reached a radius of 50$r_{\rm s} \sim 10^8$~cm.

Therefore, it is possible that the dipping is due to the change of accretion rate on the viscous timescale or is related to some instability due to thermal effects on the wind or disk. If the accretion rate is indeed several hundred times the Eddington rate, a thick disk seems more plausible, but should be tested with future simulations. 

\subsection{Optical emission and the companion}

For most broad-band ULXs, their optical spectrum is argued to be dominated by emission from the accretion disk instead of the companion star \citep{Tao2012_HoII}. In the optical band, they display a power-law like continuum that can be interpreted as emission from an irradiated accretion disk \citep[e.g.,][]{Grise2012,Cseh2013}. For NGC 247 ULX, the optical spectrum without UV data is also consistent with a power-law form with an IR excess. The SBC data rule out the power-law model, provide crucial constraints on the emission mechanism. The majority of optical emission is unlikely due to disk irradiation for NGC 247 ULX.The fitting with an irradiation model is unsuccessful especially in the UV band (Figure~\ref{fig:imcd}). The derived irradiating luminosity is one order of magnitude higher than the observed X-ray luminosity or the bolometric blackbody luminosity. The blackbody spectrum implies that we may see the companion star, which has a temperature of 19000 K and a radius of 20 $R_\sun$, consistent with those for a late O or an early B type supergiant.  Detection of the companion star means that this ULX could be a good target to search for stellar lines from the companion for a dynamical mass measurement. The presence of a second, cooler ($\sim4600$~K) blackbody component is plausible but the evidence just based on two data points in the IR band is weak. We therefore do not discuss its nature. 

\section{Conclusion}

The most important observational results from multiwavelength and multi-epoch observations of NGC 247 ULX include:
\begin{enumerate} \itemsep0pt 
\item a transition from the SSUL regime to the SUL regime -- this is the first time that a power-law emission with flux similar to the soft thermal emission has been detected in a supersoft ULX \citep[for M101 ULX, the power-law component occupies only $1/10 - 1/5$ of the total luminosity in 0.3--10 keV;][]{Urquhart2016};
\item the soft X-ray emission is highly variable on short timescales in the SSUL regime but much less variable in the SUL regime;
\item an anti-correlation between the blackbody radius and temperature, $R_{\rm bb} \propto T_{\rm bb}^{-2.8 \pm 0.3}$;
\item flux dips with a transition timescale of $\sim$200~s that cannot be explained by an increase of absorption column density; and
\item a blackbody optical spectrum with a temperature of 19000 K and a radius of 20 $R_\sun$, subject to an extinction of $E(B-V) = 0.122$.
\end{enumerate}

We interpret the observations as follows.

\begin{enumerate} \itemsep0pt
\item Supersoft and soft ULXs may be ULXs in a special accretion regime, possibly with higher accretion rates than broad-band ULXs. A standard disk explanation for the soft X-ray emission from supersoft and soft ULXs can be ruled out. An outflow model is favored, with soft X-rays originating from the photosphere of the outflow and hard X-rays being leaked emission from the inner disk via the central funnel or advected through the wind. The presence of an absorption edge near 1 keV or an optically-thin thermal plasma emission component could be signatures of outflow; Based on the \citet{Soria2016} model, the NGC 247 ULX is likely a massive stellar mass black hole accreting at $\sim 10^2$ times of the Eddington rate. Alternatively, a scenario that supersoft ULXs being standard broad-band ULXs viewed at high inclination angles can not be ruled out;  a large and variable equatorial occulter may produce the variability and dips.
\item The flux dips from NGC 247 ULX appear to be produced by changes in the photosphere of the outflow and may result from sudden increases of the accretion rate or from thermal fluctuations in the wind or disk.
\item The UV/optical spectrum if NGC 247 ULX is likely dominated by emission from the companion star, which may be an OB type supergiant.
\end{enumerate}

\acknowledgements 
We thank the anonymous referee for useful comments that have helped improve the manuscript. We also thank Yanfei Jiang, Roberto Soria, Ryan Urquhart, and Li Ji for helpful discussions, and the {\it XMM-Newton}/{\it Chandra}/{\it Swift}/{\it HST} teams for successful executions of the observations. HF acknowledges funding support from the National Natural Science Foundation of China under grant No.\ 11633003, and the Tsinghua University Initiative Scientific Research Program. PK acknowledges partial support from STScI grant HST-GO-13425.

{\it Facilities:} \facility{XMM}, \facility{Chandra}, \facility{Swift}, \facility{HST}


\end{document}